\def\year{2021}\relax

\documentclass[letterpaper]{article} 
\usepackage{subfig}
\usepackage{float}
\restylefloat{table}
\usepackage{booktabs,caption}
\usepackage[flushleft]{threeparttable}
\usepackage{csquotes}
\usepackage{amsmath}
\usepackage{hyperref}
\usepackage{xltabular}
\usepackage{xcolor}
\usepackage{aaai21}  
\usepackage{times}  
\usepackage{helvet} 
\usepackage{courier}  
\usepackage{graphicx} 
\usepackage{natbib}  
\usepackage{caption} 
\usepackage{array}

\title{What Are You Anxious About? Examining Subjects of Anxiety during the COVID-19 Pandemic}
\author {
     Lucia L. Chen,\textsuperscript{\rm 1}
    Steven R. Wilson, \textsuperscript{\rm 2}
    Sophie Lohmann \textsuperscript{\rm 3} 
    Daniela V. Negraia \textsuperscript{\rm 3,4}\\
}
\affiliations {
    \textsuperscript{\rm 1} Department of Health Policy, Stanford University, United States \\
    \textsuperscript{\rm 2} School of Engineering and Computer Science, Oakland University, United States \\
    \textsuperscript{\rm 3} Max Planck Institute for Demographic Research \\
    \textsuperscript{\rm 4} Department of Sociology, Oxford University, United Kingdom\\
    lucia.chen@stanford.edu, stevenwilson@oakland.edu, lohmann@demogr.mpg.de, negraia@demogr.mpg.de
    
}

\begin{document}

\maketitle

\begin{abstract}

 COVID-19 poses disproportionate mental health consequences to the public during different phases of the pandemic. We use a computational approach to capture the specific aspects that trigger an online community's anxiety about the pandemic and investigate how these aspects change over time. First, we identified nine subjects of anxiety (SOAs) in a sample of Reddit posts ($N$=86) from r/COVID19\_support using thematic analysis. Then, we quantified Reddit users' anxiety by training algorithms on a manually annotated sample ($N$=793) to automatically label the SOAs in a larger chronological sample ($N$=6,535). The nine SOAs align with items in various recently developed pandemic anxiety measurement scales. We observed that Reddit users' concerns about health risks remained high in the first eight months of the pandemic. These concerns diminished dramatically despite the surge of cases occurring later. In general, users' language disclosing the SOAs became less intense as the pandemic progressed. However, worries about mental health and the future increased steadily throughout the period covered in this study. People also tended to use more intense language to describe mental health concerns than health risks or death concerns. Our results suggest that this online group's mental health condition does not necessarily improve despite COVID-19 gradually weakening as a health threat due to appropriate countermeasures. Our system lays the groundwork for population health and epidemiology scholars to examine aspects that provoke pandemic anxiety in a timely fashion.
\end{abstract}

\section{Introduction}
The novel Coronavirus Disease (COVID-19) pandemic has created a global crisis. Unlike other recent pandemics (e.g., H1N1 or type-A influenza), the COVID-19 pandemic has resulted in strict and extensive lockdown measures for large swathes of the global population, such as complete lockdowns, 14-day quarantine periods, closing of national borders, and disruption of international travel. A series of life interruptions resulting from the COVID-19 pandemic are related to heightened anxiety and depression in many people \citep{smith2020correlates}. In a 2020 study, over 80\% of respondents reported that their day-to-day thoughts were occupied by topics related to the COVID-19 pandemic \citep{roy2020study}. Sleep difficulties, extreme anxiety about becoming infected with COVID-19, and distress caused by information from social media were reported by 12.5\%, 37.8\%, and 36.4\% of participants, respectively \citep{roy2020study}. 


Anxiety is conceptualized as a multi-system response to perceived risks, experienced as a feeling of unease, worry, or fear \citep{wilkinson2001anxiety}. Experiencing occasional anxiety about a situation is a normal part of life. However, prolonged or frequent anxiety may lead to general anxiety disorders, heightened depressive symptoms, reduction of sleep quality \citep{huang2020generalized}, and other mental health risks. Psychological studies often use scales to assess levels of anxiety. Studies that examine COVID-19 anxiety mainly fall into two categories: those that treat anxiety as a clinical entity with a focus on the symptoms \citep{ahorsu2020fear} and those that examine the specific aspects of the pandemic that trigger anxiety \citep{mcelroy2020demographic, taylor2020development}. Although methodologies that use a survey approach with scale measurements can provide detailed descriptions of individual respondents, their implementation is expensive and time-consuming. 

In an effort to collect data in a timely and cost-effective manner, researchers have been exploring avenues to approximate the anxiety of the public using proxy signals in social media records \citep{guntuku2019twitter}. Our study does not aim to replace the measurement using scales because some of the behavioral, affective, and cognitive characteristics measured with their items may rarely be reflected in the social media text. Rather, our automatic system may provide a proxy value for the scientific and public health community to understand better how the public responds to large-scale health threats promptly.

Our objective is to use data from Reddit to understand the aspects that provoke people's anxiety during early phases of the pandemic and how these aspects have changed as the pandemic has developed. Reddit is an online platform in which a network of communities is organized around users' interests. Reddit users engage with these communities, also known as subreddits, by submitting text posts, links, images and comments. In line with the platform's conventions, for the remainder of this paper we will use the prefix ``r/'' to denote specific subreddit communities (e.g., r/COVID19\_support; a subreddit that is specifically for people sharing advice and coping mechanisms for COVID-19). Focusing on a sample of Reddit users who have posted in r/COVID19\_support, we propose the following research questions (RQs):

\begin{itemize}
    
    \item \textbf{RQ1:} \textit{What are the subjects of anxiety (SOAs) expressed in r/COVID-19 support? How do the SOAs change over time?}

   \item \textbf{RQ2:} \textit{How does the intensity of anxiety-related language change over time? }
   
   \item \textbf{RQ3:}  \textit{Which SOAs were described by users with high anxiety-related language?}

\end{itemize}

To answer these questions, we first conduct a thematic analysis to identify the subjects of anxiety (SOAs) mentioned in posts from r/COVID19\_support. Thematic analysis is a qualitative data analysis method for identifying topics and concepts that occur repeatedly in the data. Following the thematic analysis, we identified nine SOAs in the Reddit posts. Then we define detailed annotation guidelines for identifying these SOAs in 793 posts (12\% of the data) extracted from r/COVID19\_support. Each post could contain more than one SOA. At the time we developed our annotation guidelines, there were no published guidelines on measuring COVID-19-related anxiety. Later on, we found that the SOAs in our annotation guidelines overlapped with items in the recently developed COVID-19 anxiety measurement scales (C-19ASS) (18 items) \citep{nikvcevic2020covid}, COVID-19 Concern Questionnaire (CCQ)(6-items), COVID-19 Experiences Questionnaire (CEQ) (14 items) \citep{conway2020social} and the COVID Stress Scales (33-items) \citep{taylor2020development}, providing further support for the widespread existence SOAs which we identified (see Appendix table \ref{tab:SOWanno}). 

Next, we turn our attention to the intensity of the language used to describe each SOA. In the example ``I am super anxious about...'',  ``super anxious'' indicates that the author has a high level of anxiety towards the SOA. Importantly, however, the level of anxiety we annotated only reflects the author's anxiety towards the specific topic that they wrote about in that particular post. We do not have information on their anxiety over time or about other topics and our analysis therefore does not reflect a holistic, clinical definition of anxiety. Finally, using the annotated posts, we train multi-label classifiers based on both support vector machines (SVMs) and deep learning architectures. We then use the SVM classifier to annotate the remaining posts in r/COVID19\_support ($N = 5747$), then analyze the trends and patterns of the machine-annotated labels. 

\subsection{Contributions}
This study makes several methodological and substantive contributions to the study of mental health using social media data. First, by examining data from an online social media platform where users can maintain their anonymity, we had the opportunity to study naturalistic reports about people's personal life experiences during a global pandemic. Second, we developed an annotation scheme for aspects that trigger anxiety among social media users. Third, we developed a machine-annotated pipeline to identify these aspects, adding new insights to the extant knowledge derived from survey studies \citep{nikvcevic2020covid, conway2020social, taylor2020development}. A machine-annotated pipeline allows us to track how these aspects change across time. Fourth, by evaluating our results on new incoming data, we highlight the importance for researchers to update their knowledge about social media content while constructing models to infer mental health status. Our approach can help public health practitioners understand how the public responds and adjusts to a pandemic in real-time.  

\section{Prior Work}
\subsection{Disclosure of anxiety on social media}

Seeking support can be helpful for people with anxiety \citep{roohafza2014s}. Support is especially important during a pandemic when stay-at-home orders curtail the possibility of in-person consultations between practitioners and clients. However, people often find it difficult to do so for many reasons. Some people may hesitate to seek professional help and doubt the efficacy of medical resources because they attribute mental illnesses to personal weakness rather than illness \citep{yap2011influence}. 

Fortunately, help-seeking behavior can be facilitated by various forms of social media. Social media platforms such as Reddit that encourage anonymity allow people to talk about their mental health issues without fear of being judged or identified. During the COVID-19 pandemic, social media platforms, among other services, connected people who were struggling with social isolation. These platforms facilitate the sharing of useful information during user interactions which can help them to cope with anxiety \citep{wiederhold2020using}. 

Audiences in support communities often respond soon after a post is published, providing timely support to those seeking help \citep{pfefferbaum2020mental}. Additionally, helping other people deal with their struggles is not only beneficial for the recipient but also for the support provider \citep{brown2003providing}. Prior work shows that online support communities provide mental health benefits to people seeking help from the platform \citep{bargh2002can}. The linguistic attributes of social media content also reflect users' depressive symptoms \citep{chancellor2020methods}, addiction \citep{murnane2014unraveling} and other mental health concerns. 



During a pandemic, social media users often describe their struggles in plain text. Anxiety is often expressed in social media posts and recent studies have taken steps to document it. For example, \citet{jones2020not} annotated anxiety expressed in social media posts according to anxiety words measured by the Linguistic Inquiry and Word Count (LIWC) program.  In LIWC, words that reflect anxiety include ``afraid'', ``scared'',  ``worried'' and 111 other words \citep{pennebaker2015development}. \citet{shen2017detecting} used textual features (e.g., topic modeling, word embeddings) to distinguish posts in anxiety-related Reddit communities from other Reddit communities. Recently, researchers developed tools that facilitate early recognition of hot-spots of declining mental health by detecting anxiety language from tweets \citep{guntuku2020tracking}. During COVID-19, computational approaches revealed that language related to ``economic stress'', ``isolation'' and ``home'' increased significantly among many Reddit mental health support groups (e.g., r/addiction, r/alcoholism, r/adhd, r/anxiety, r/autism, r/BipolarReddit, r/bpd and r/depression) \citep{low2020natural}.


\section{Data} \label{sec:data}

Social media platforms feature different functionalities, and these functionalities influence users' motivation to post on the platform. Reddit is not only a platform for sharing news and interests; it also comprises many mental health support communities that allow people to focus on sharing, talking, and fostering group connections \citep{chen2021monitoring}. Reddit is more suitable for identifying themes in discussions and conversations than Facebook and Twitter. The anonymous nature of Reddit allows people to talk about their mental health issues without the fear of being stigmatized. In this study, we focus on a community (``r/COVID19\_support'') for people interested in support navigating the COVID-19 pandemic. ``r/COVID19\_support'' is the only subreddit dedicated to this specific purpose (news sharing was forbidden), which was created on 12th February 2020 and had 29.9k users by April 2021. 

We collected the data from the Reddit API using The Python Reddit API Wrapper (PRAW) \citep{praw}. The data collection process took place between April 2020 and April 2021. Our dataset contains all available posts ($N$=6,540, 1.92 posts per user) between the inception of the ``r/COVID19\_support'' community (12th February 2020) and the end period of the present analysis (28th April 2021). This dataset covers 13 months of the COVID-19 pandemic. Posts that were flagged as ``deleted'' or ``removed'' by the post authors were removed from the sample. 

\subsection{Ethics and Privacy} 

The data used in this study were publicly available when collected from (\hyperlink{https://www.reddit.com/}{Reddit}). Data were secured on firewalled servers to ensure data protection, and researchers could download the data only on local machines. Researchers were not allowed to share data and had no interaction with the users. To protect the privacy and anonymity of the users in our dataset, we paraphrased the quotes published in this paper. The annotated dataset will not be released but the model pipeline is published on  \href{https://github.com/luciasalar/pandemic_anxiety}{Github} and annotation guidelines will be shared. 

\section{Methodology}


\subsection{Thematic Analysis for Subjects of Anxiety}

\textit{Identifying Users' Subjects of Anxiety (SOA).} We adopted thematic analysis \citep{braun2006using} to answer our first research question, ``What are the SOAs expressed in r/COVID-19\_support?''.  We refer to studies with a focus on qualitative analysis to determine the number of posts we need to analyze. In general, interview-based qualitative analysis researchers suggest that large interview studies often contain 50-60 interviews \citep{britten1995qualitative}. In our case, there are no existing guidelines on the number of participants for the analysis of social media text. Recently conducted content analyses on Reddit posts often included 100-200 annotation documents \citep{maxwell2020short}. In this work, we randomly selected 90 posts from r/COVID-19\_support between February and August 2020 (around 10 posts per month). Posts without content and those that only contained URLs or images were removed, resulting in 86 posts for annotation, with an average word count of 161. Since we have roughly 10 posts per month for the initial thematic analysis for subjects of anxiety, it is possible that some users’ sources of anxiety may have been missed.

Initially, the first author read and re-read a post to identify potential SOAs. The annotation includes a brief description of various SOAs identified in the text. Then the first author summarized the SOAs into several main groups and subgroups (see Figure \ref{fig:SOAgraph}). The groups were then forwarded to the third and fourth authors. The second level of analysis involved both the third and fourth authors reviewing some of these initial coding examples and the SOA groups. They particularly considered how to retain the initial codes' diversity while producing overarching elements and sub-groups. Disagreement of the codes were resolved through discussion. 

\subsection{Manual Annotation of Subjects of Anxiety} 
After we identified the SOAs with thematic analysis, three authors of the present study compiled the annotation guidelines (\url{https://bit.ly/3mUYMGR}). Next, we used stratified random sampling to select 100 posts from each month (between February and Oct 2020). We pooled the data from February and March 2020 because there were only 85  posts in February 2020. After removing posts that did not contain content or contain hyperlinks only, we obtained 793 posts for the manual annotation task. Two annotators cross-annotated each post, and a third annotator resolved any annotation conflicts. Three of the authors and a student helper collaborated on this task, and the annotation agreement (Cronbach's alpha) was 0.82.

\subsection{Identifying Anxiety-related Language}
\textit{Annotating the intensity of users' anxiety-related language.} We further annotated the intensity of users' anxiety-related language present in each post. We defined three intensity levels: Level 0: user did not express anxiety or fear, for example, ``I came across this article from New York Times. Thoughts on accuracy?'' Level 1: user expressed some level of anxiety or fear but no urgency or desperation, for example, ``I found that I’m always thinking about COVID related things. It seems so normal to me to feel this way now.'' Level 2: user expressed extreme anxiety or fear, such as ``I am so, so, so scared'', or ``My anxiety is going through the roof''. The annotation agreement for the anxiety-related language was Cronbach's alpha = 0.72. 

\textit{Rater Credibility and Expertise} Three out of the four annotators are authors of this paper who have psychological expertise and research experience in both quantitative and qualitative studies of mental health. The fourth annotator was a Master's-level student in demography who received training from the study authors prior to completing the annotation task.

\subsection{Machine Annotation}
 We used the annotated data ($N = 793$) to construct a multilabel support vector machine (SVM) classifier, composed of an ensemble of binary classifiers, to automatically identify the SOAs present in a given Reddit post. 
 
 \textit{Combining Labels}. We found that the posts with a language intensity level 0 (no signs of anxiety) only accounted for about 10\% of the data ($N=80$).  80\% ($N = 64$) of the level 0 posts did not mention any SOA. This included posts that only had a title but no content, or only emoji in the content. Level 0 posts were mainly noise for the anxiety language analysis task, but we did not construct a separate classifier to remove level 0 posts for this analysis because the number of posts was too small for a training sample. Instead, we aggregated level 0 with level 1 (posts expressing mild anxiety), then built a binary classifier for anxiety-related language intensity, i.e., the classifier is trained to predict whether a given post's intensity should be categorized as level 2 (posts expressing extreme anxiety of fear) or not.

\textit{Feature Selection}. We adopted features (e.g., sentiment and topics) that are commonly used in classifying mental illness symptoms \citep{chancellor2020methods, chen2020examining}. Features involved in the SOAs and language intensity including: 1) N-gram word representation (tf-idf count vector), where $n\in N := \{1, 2, 3\}$. The preprocessing of n\_grams included removing stopwords, lowercasing, ignoring terms that appear in more than 50\% of the documents or less than 0.25\% of the documents. 2) Sentiment. We used the Valence Aware Dictionary and Sentiment Reasoner (VADER), which is a rule-based sentiment analysis tool specifically for detecting sentiment expressed in social media text \citep{hutto2014vader}, to generate a positive, negative and neutral score for each text. 3) 15 topics extracted with LDA topic modeling. We experimented with 10, 15 and 20 topics, and found that 15 topics yielded the highest coherence score(U\_MASS measure \citep{mimno2011optimizing}). Topic coherence measures the degree of semantic similarity between high scoring words in the topic. 4) The Flesch reading ease score. The reading ease score indicates the readability of a text by computing the average length of the sentences and the average number of syllables per word. A lower score indicates more complicated and long words are used in the text, making it more difficult for readers to process the text \citep{coleman1975computer}. 

We adopt a \textit{``one vs rest''} approach to train an ensemble of nine binary classifiers. In the main model, we split the data into a train (80\%, N = 634) and test set (20\%, N = 159) in a stratified fashion. Stratified five-fold cross-validation in the training set was used to optimize the hyperparameters in the model training and a grid search of hyperparameters was carried out for the classification algorithms. For the baseline models, we constructed dummy classifiers that generate predictions by respecting the training set's class distribution (See Sklearn DummyClassifer) \citep{scikit-learn}.

\subsection{Model Validation on Chronological Data}
We assumed that the aspects that provoke the public's anxiety would change over time, therefore, it's important to validate that models trained on historic data perform well on subsequent data. In the main model, we used models trained on a dataset from Feb 2020 - Oct 2020 to predict data collected after Oct 2020. We employed two approaches to examine the model validation on chronological data. 1) ``Last month holdout''. We constructed a set of models trained on data from Feb 2020 - Sept 2020. Stratified five-fold cross-validation and a grid search of hyperparameters was carried out for the classification algorithms. Then, we tested the models on a holdout set in Oct 2020. 
 2) ``Human evaluation on Incoming data''. We were also aware that new SOAs might have emerged after we completed the thematic analysis for SOAs in Aug 2020. To examine the model's validity on new data, especially data after 2020, we conducted human evaluation on 50 posts randomly selected during 1st Mar 2021 and 30th Apr 2021. The first author conducted another round of thematic analysis and SOA annotation on the 50 posts.

\subsection{Comparison to Deep Learning Models}
Some works on using social media data to infer mental illnesses symptoms adopted deep learning techniques \citep{shen2018cross}. Therefore, we also experiment with a fine-tuning DistilBERT model for our task \citep{sanh2019distilbert}. 
For this task, the annotated dataset is divided into train, validation, and test sets with a proportion of 0.7:0.1:0.2. We used Huggingface Transformers \citep{wolf2019huggingface} combined with hyperparameter search using the Ray Tune Python library \citep{liaw2018tune}. 


\section{Results}

\subsection{Summary of Statistics}
We collected 6535 posts from January 2020 to April 2021 from r/COVID19\_support. Posts that were removed or deleted by the author by the time we collected the data were not included in the dataset. Table \ref{tab:stats} shows the number of posts each month. We found a dramatic surge of posts in March 2020 followed by a tapering off of posts over the following months. Users may have become less enthusiastic about COVID-19 discussions or moved these discussions to other forums.

\begin{table}
\fontsize{7}{7}\selectfont
\setlength\tabcolsep{1.5pt}
\centering
\begin{tabular}{l|rrrrrrrrrrr|rrrr}
\hline  & \\[-1.5ex]
 year      & \multicolumn{11}{l}{2020}                                                                      & \multicolumn{4}{l}{2021}  \\
 month      & 2    & 3    & 4    & 5    & 6    & 7    & 8    & 9    & 10   & 11   & 12   & 1    & 2    & 3    & 4    \\
       \hline  & \\[-1.5ex]
posts  & 12   & 1023 & 745  & 481  & 333  & 538  & 417  & 400  & 409  & 409  & 470  & 466  & 329  & 206  & 297  \\
p/user & 1.09 & 1.56 & 1.42 & 1.38 & 1.44 & 1.41 & 1.40 & 1.39 & 1.31 & 1.29 & 1.38 & 1.49 & 1.53 & 1.44 & 1.47\\
\hline
\end{tabular}
\caption{Statistics for collected posts. Posts: number of posts collected in r/COVID19\_support in the corresponding month. p/user: number of posts per user.}
\label{tab:stats}
\end{table}

\subsection{Identifying the Subjects of Anxiety}



We identified nine groups of subjects of anxiety (SOA) in the thematic analysis. The authors and student helper co-annotated 793 posts according to the annotation guideline for Subjects of Anxiety (SOA). Among these, 645 contained at least one SOA, 148 (18\%) did not match any of our annotation topics, and 379 (50\%) posts contained more than one SOA. All of the SOAs could be directed either towards self or towards friends or family members.

The nine groups of SOA include worry, fear, or anxiety ($N$ indicates the number of posts labeled as positive in the corresponding category): 1) about COVID-19 leading to financial difficulties or stagnation in one’s career/school progress (Career/Finance, $N=98$); 2) of being infected with COVID-19 in general, including fears that others (e.g., friends or family members) will get infected  (Health Risks, $N=382$); 3) of being infected with COVID-19 because of going to work in an environment where protection measures would be insufficient, or physical distancing would not be possible (Work Risks, $N=52$); 4) that strangers/friends/family members did not follow guidelines for preventative measures (Disrespecting Health-Related Guidelines, $N=164$); 5) of traveling using public transportation (Travel Risks, $N=48$); 6) of oneself or loved ones dying due to COVID-19, or becoming critically ill (Death anxiety, $N=66$); 7) about when to quarantine if a person is tested positive and general lockdown for the public, what the guidelines for quarantine/social distancing are, when quarantine/mandated social distancing will finish, or conflicts with familiar members in a quarantine/social distancing situation, or struggling with quarantine/social distancing and not seeing loved ones (COVID-19 Restrictions, $N=164$). The final two subjects captured descriptions 8) of deteriorating mental health, or worry that one's mental health may worsen (Mental Health, $N=188$); 9) of losing hope for the future (Future, $N=98$). Below are the categories and examples of our annotation (these examples are paraphrased to protect user privacy).

\begin{figure*}
\centering
    \includegraphics[scale = 0.48]{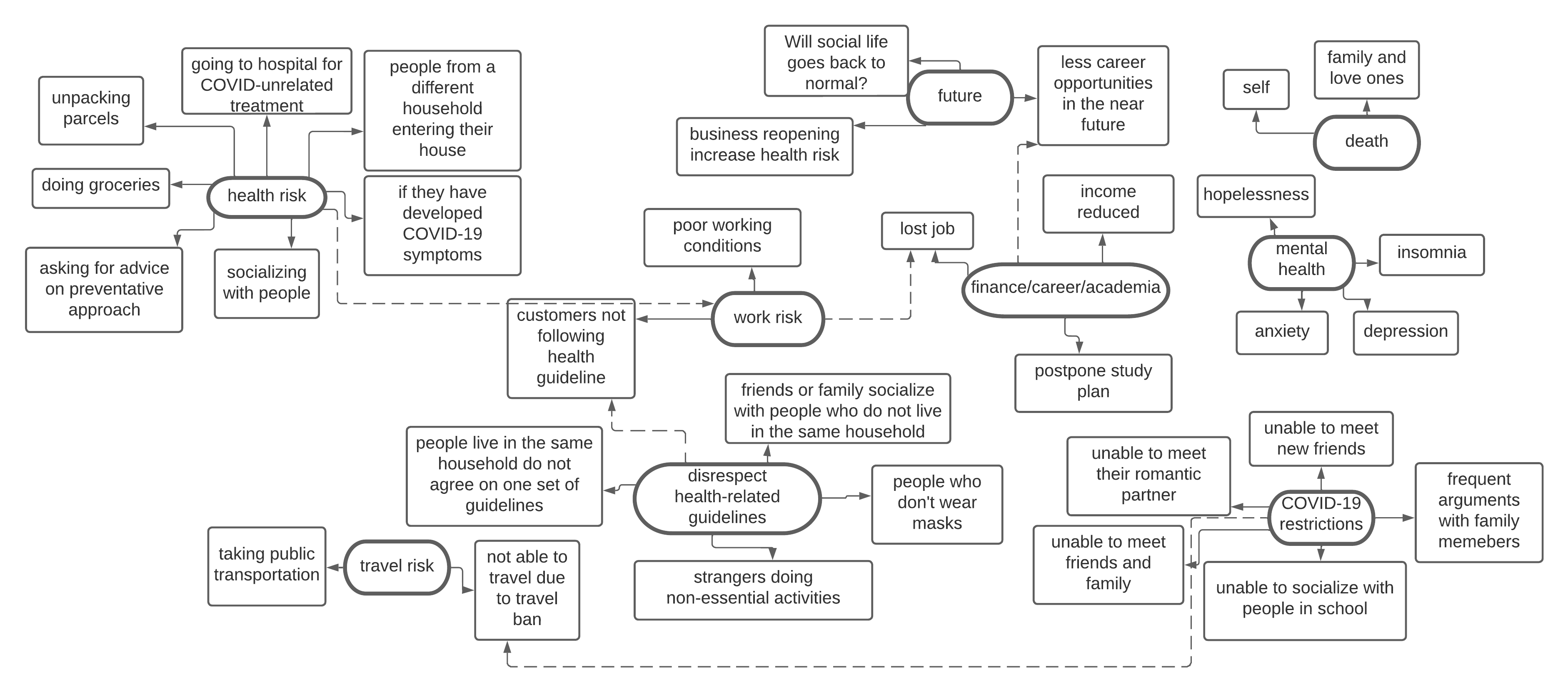}\\

    \caption{Subgroups for manually annotated SOAs in Thematic Analysis. This graph shows the subgroups in each SOA. Dotted lines indicate SOAs that have overlapping subgroups and SOAs which are subcategories of others.} 
    \label{fig:SOAgraph}
\end{figure*}

\subsubsection{Worry about Health Risk.}

At the beginning of the pandemic, due to the lack of knowledge about how the virus spread, users were worried about infection risk partially because they did not know the appropriate preventative measures. We identified several sub-groups under this SOA. Users often sought information about whether they were using the right preventative approach (see Figure \ref{fig:SOAgraph}), such as whether they should go to the hospital for a COVID-19 unrelated treatment, or whether they could be infected with COVID-19 by unpacking parcels, for example:

\begin{quote}
\textit{Can I get the virus by unpacking boxes if they are shipped today and come Tuesday?}
\end{quote}

Some users also listed symptoms and asked if those were symptoms of COVID-19. We observed that people with preexisting health conditions were more likely to have this concern. 

\blockquote{\textit{My normal body temperature is 96, but I have been up to 99.5 on and off for a week. I have mild asthma. I have sleep apnea and I have a slight dry cough. Now I'm afraid I might have it.}}

\subsubsection{Worry about Work Risk.}
Worry about work risk is a subgroup of health risk (see Figure \ref{fig:SOAgraph}) (all work risk posts were also labeled as health risk). People worried about being infected at work and were often struggling with multiple problems, such as customers or patients disrespecting health-related guidelines. Users often revealed that they or their family members were essential workers and that they worried about being infected in the working environment, for example:

\begin{quote}
\textit{My mom is a lung doctor and she was working with patients in the ICU. I'm really scared.}
\end{quote}

 Some users were unhappy about poor working conditions. They were also worried about losing their jobs or missing job opportunities if they did not continue to work despite potential risks.

\blockquote{\textit{I saw an ad on Facebook saying they're looking for employees at a small grocery store nearby, and I wanted to apply. But I'm a little nervous about exposing myself like this. I'm currently unemployed, and my savings will only last a few months.}}

\subsubsection{Worry about COVID-19 restrictions.}
As the pandemic progressed, around April 2020, issues related to COVID-19 restrictions (e.g., quarantine, social distancing, and lockdowns) emerged as a major concern. Reddit users expressed feelings of distress and loss about not being able to meet their loved ones for months:

\begin{quote}
\textit{It's been more than two months of social isolation, and all I want to do now is be physically with my partner. I don't know when I'll be able to see her again because she's immunocompromised. That's a bummer...}
\end{quote}

At the same time, restrictions related to the COVID-19 pandemic have forced people to spend more time together at home. In some cases, users' posts suggested that they were experiencing greater conflict with loved ones than before the pandemic. We found these posts occurred more often at the beginning of the pandemic.

\begin{quote}
\textit{I'm really frustrated because the pandemic is causing so much conflict between me and my wife. My wife is getting incredibly angry at me with little or no provocation. Little things are blown up into massive arguments.}
\end{quote}

\subsubsection{Worry about Career/Finance.}
As quarantine and social distancing were needed to stop the spread of the virus, many users, especially essential workers, mentioned that they were worried about financial or career situations. Some users mentioned that they postponed their study plans due to a lack of motivation to carry out their studies online or financial difficulties. We combined worry about academic progress with worry about one's career into one category because the academic progress is closely related to one's career, for example:

\begin{quote}
\textit{I had dreams of completing my college degree, which I would have probably finished in 1.5 years. It will most likely take at least another two years to finish now. I had hoped to teach music or start my own business in the near future, but that is no longer viable. }
\end{quote}

Many users reported struggling with financial situations because they were unable to find a part-time job.

\begin{quote}
\textit{I'll be starting college soon, and I'm as broke as everyone else. I really need this money if I'm going to survive.}
\end{quote}

\subsubsection{Worry about Disrespecting Health-related guidelines.}
We found that many users developed their own set of health-related guidelines. Users reported feeling uncomfortable or distressed that friends, family, or strangers did not respect their health-related guidelines. For example, they disagreed with others that they lived with over whether they should or should not visit or socialize with non-household members.  

\begin{quote}
\textit{My parents are going on walks with friends (across the street but still make me nervous).}
\end{quote}

\begin{quote}
\textit{I've set many boundaries with my roommate. First was no visitors at all. She got around it by smuggling someone in. Then visitors only in her room, with her sanitizing everything afterward. She broke that blatantly.}
\end{quote}

\subsubsection{Worry about Traveling and Public Transportation.}
Starting in March 2020, many countries imposed travel restrictions or travel bans in an attempt to reduce the spread of the virus. Users posted that these travel restrictions were a source of anxiety especially for those who couldn't visit their partners or family. For example:

\begin{quote}
\textit{I need to return home sometime this year, but I'm almost sure I missed my window of opportunity. Am I fucked for the rest of the year? How can I go home? :( Please help.}
\end{quote}

Users posted about being worried to use public transportation in general, and particularly about air travel:

\begin{quote}
\textit{I am going to fly out to visit family for a week in the first week of July and I am extremely anxious about it.}
\end{quote}

\subsubsection{Worry about Death.}
At the beginning of the pandemic, most of the concerns about the pandemic centered around infection risks or feeling anxious about elderly family members becoming severely ill, especially when the number of reported cases was surging, for example:

\blockquote{\textit{Not only am I terrified of dying, but I'm also worried about my parents and grandfathers. Every day on social media, I see the number of deaths and mourn because so many people have lost loved ones...}}

\subsubsection{Worry about the Future.}
Users also expressed anxiety about the immediate and future economic impacts of the COVID-19 pandemic. These concerns were related to career opportunities, being furloughed or laid off and interruption of school, for example:

\blockquote{\textit{I'm a student at college. I just received acceptance to a university that said that all classes would be held online for the full semester. My wife will begin her new career from home as well. Everything that used to be nice in my life makes me miserable. }}

\subsubsection{Worry about Mental Health.}
In October 2020, about seven months after the start of the pandemic, we observed that a higher share of users expressed hopelessness, and concerns about mental health deterioration becoming more severe. Concerns about mental health remained high until April 2021, when we finished the data collection for this study. 

\blockquote{\textit{I don't know how much longer I can continue. Since March, I've been in and out of severe anxiety and depression episodes when this began. I have health anxiety and this is my worst nightmare...}}

\subsection{Classifying Subjects of Anxiety}

We constructed an ensemble model comprising nine binary classifiers for the SOAs using SVM and a fine-tuned DistilBERT model (main models). Table \ref{tab:model} shows the results of the SOA classifiers and language intensity classifier. In the SVM main model, the per-class F1 scores were above 0.60 for half of the classes. However, categories with extremely imbalanced negative to positive classes, such as work (149: 10), death (146:13), travel (149:10), and future (139:20) had poorer model performance but still surpassed the baseline dummy classifier. The DistilBERT model performed better in the health risk, travel and future categories but performed poorer than the SVM in the rest of the categories. We also compared the main models' performance with logistic regression and random forest \ref{tab:model}. The random forest algorithm showed the poorest performance among all the algorithms we tried, whereas the logistic regression models performed worse than SVM and neural network models.

\textit{Chronological data analysis}. Since the SVM model had a more satisfactory result than DistilBERT in the majority of the categories, we focus on evaluating the SVM model in the chronological data analysis. Results from the ``last month holdout (SVM)'' models were similar to the SVM main models, except for ``disrespecting health-related guidelines'' and ``death''. Our results suggest that although the SOA annotation guideline was defined according to the thematic analysis conducted on data between Feb 2020 and Aug 2020, the automatic SOA annotation models perform satisfactorily on data in Sept and Oct 2020. However, models trained on infrequently occurring variables are less stable. 

\begin{table*}
\centering
\caption{Model Performance. P:precision, R:recall and F1-score of positive class on subjects of anxiety. Macro average is reported for language intensity. ratio: ratio of positive class in test set. finance: career or finance, restrict: COVID-19 restrictions, health: health risks (general), guide: disrespecting health-related guidelines, work: work risks, mental: mental health, intensity: anxiety-related language intensity}
\tiny
\setlength\tabcolsep{4pt}
\centering
\begin{tabular}{r|r|rrr|rrr|rrr|rrr|rrr|rrr||r|rrr}
\hline
& &\multicolumn{3}{c}{SVM} & \multicolumn{3}{c}{DistilBERT}  & \multicolumn{3}{c}{Logistic Regression} & \multicolumn{3}{c}{Random Forest} &

\multicolumn{3}{c}{baseline} & \multicolumn{3}{c} {last month holdout}&  \multicolumn{4}{c}{2021 Data}\\
variable  & ratio       & P & R & F1  & P & R & F1 & P & R & F1 & P & R & F1 &  P & R & F1 & P & R & F1 & ratio & P & R & F1  \\ 
\hline
intensity & 36\% & 0.70      & 0.71   & 0.71       & 0.64 & 0.68 & 0.66  & 0.70        & 0.67        & 0.68        & 0.70      & 0.62      & 0.61  & 0.52 & 0.52 &   0.52 &  0.79 & 0.73 &   0.74  &  high:33\% &  0.76 & 0.75 & 0.75 \\ \hline
finance   & 13\%    & 0.75      & 0.45   & 0.56        &0.57 & 0.85& \textbf{0.68}  & 0.88        & 0.35        & 0.50        & 0         & 0         & 0  &  0.08 & 0.1&    0.09& 0.67 & 0.75 &  0.71      & 4\% & 0.30 & 0.50 & 0.40 \\
restrict  & 23\%    & 0.68      & 0.55   & \textbf{0.61}        & 0.63 &0.5 & 0.56 &0.67        & 0.42        & 0.52        & 0.90      & 0.24      & 0.38 &  0.26& 0.29 &  0.28 & 0.64 & 0.55 &  0.59    & 10\%  & 0.57 & 0.80 & 0.67    \\
health  & 48\%  & 0.72      & 0.78  & 0.75       & 0.77& 0.85 & \textbf{0.80}  &0.71&0.77& 0.74&0.70&0.74& 0.72   & 0.46 &  0.39&    0.42 &  0.82 & 0.74 &   0.78  & 20\% & 0.40 & 0.60 & 0.48 \\ 
guide    & 21\% & 0.79      & 0.45  & \textbf{0.58 }    & 1& 0.06 & 0.11  & 0.70        & 0.48        & 0.57        & 1.0       & 0.06      & 0.11   & 0.22& 0.24& 0.23  & 0.55& 0.34&   0.41        &  4\% & 0    & 0    & 0 \\
work    & 6\% & 0.50      & 0.30   & \textbf{0.38}     & 0.07& 0.83 & 0.14   & 0.50        & 0.10        & 0.17        & 0         & 0         & 0  & 0.09 & 0.1 &   0.09  & 0.50 & 0.50 &  0.50    &  2\%    & 0    & 0    & 0   \\
mental   & 22\%  & 0.72      & 0.66   & \textbf{0.69}     & 0.56& 0.68& 0.62 & 0.72        & 0.51        & 0.60        & 0.78      & 0.20      & 0.30   & 0.16 & 0.17&     0.16  & 0.65 & 0.65 & 0.65    & 12\% & 0.36 & 0.67 & 0.47 \\
death       & 8\%       & 0.42      & 0.39   & \textbf{0.40}     & 0 & 0& 0  & 0           & 0           & 0           & 0         & 0       & 0.22 & 0.15& 0.18     & 0 & 0 & 0& 0        & 2\%  & 0    & 0    & 0  \\
travel     & 6\%    & 0.50      & 0.10   & 0.17     & 0.6& 0.6& \textbf{0.60}  & 1.00        & 0.10        & 0.18        & 0         & 0         & 0  & 0 & 0 & 0   &  1.00 & 0.40 & 0.57      &  8\%  &  0    &  0    & 0   \\
future      & 13\%        & 0.55      & 0.30   & 0.39    & 0.83 &0.5 & \textbf{0.63}  & 0.80        & 0.20        & 0.32        & 1.00      & 0.05      & 0.10 & 0.08& 0.1 & 0.09 & 0.57 & 0.33 & 0.32  & 12\% & 0.25 & 0.17 & 0.20 \\
\hline
\end{tabular}

\label{tab:model}
\end{table*}

\subsection{Model Validation in 2021} \label{modelval}
To examine the model performance on data from 2021, we conducted a thematic analysis on data in March and April 2021. We found several new SOAs that were related to vaccines. Users were worried about vaccine delay, low vaccination rates, vaccine side effects, and vaccine efficacy. We also identified new sub-groups in the SOAs. For example, for health risks, users were worried that it would be unsafe to socialize or travel even after vaccination. Other infection risks included COVID reinfection and vaccine breakthrough infection. Table \ref{tab:model} shows the result evaluation for 2021 data. The performance of language intensity remains similar to data in 2020. However, classifiers with fewer positive class training data (e.g., travel, death) were again not performing well in the new dataset. Most importantly, our classifiers of course were not able to identify SOAs that had not been represented in our codings (e.g., vaccine-related worries). Although our classification results show a decreasing trend of many SOAs after Feb 2021, this does not necessarily mean that people worried less. Instead, people may have shifted to worries about the vaccine.

\subsection{How did Subjects of Anxiety Change Over Time?}
\begin{figure}
\centering
    \includegraphics[scale = 0.40]{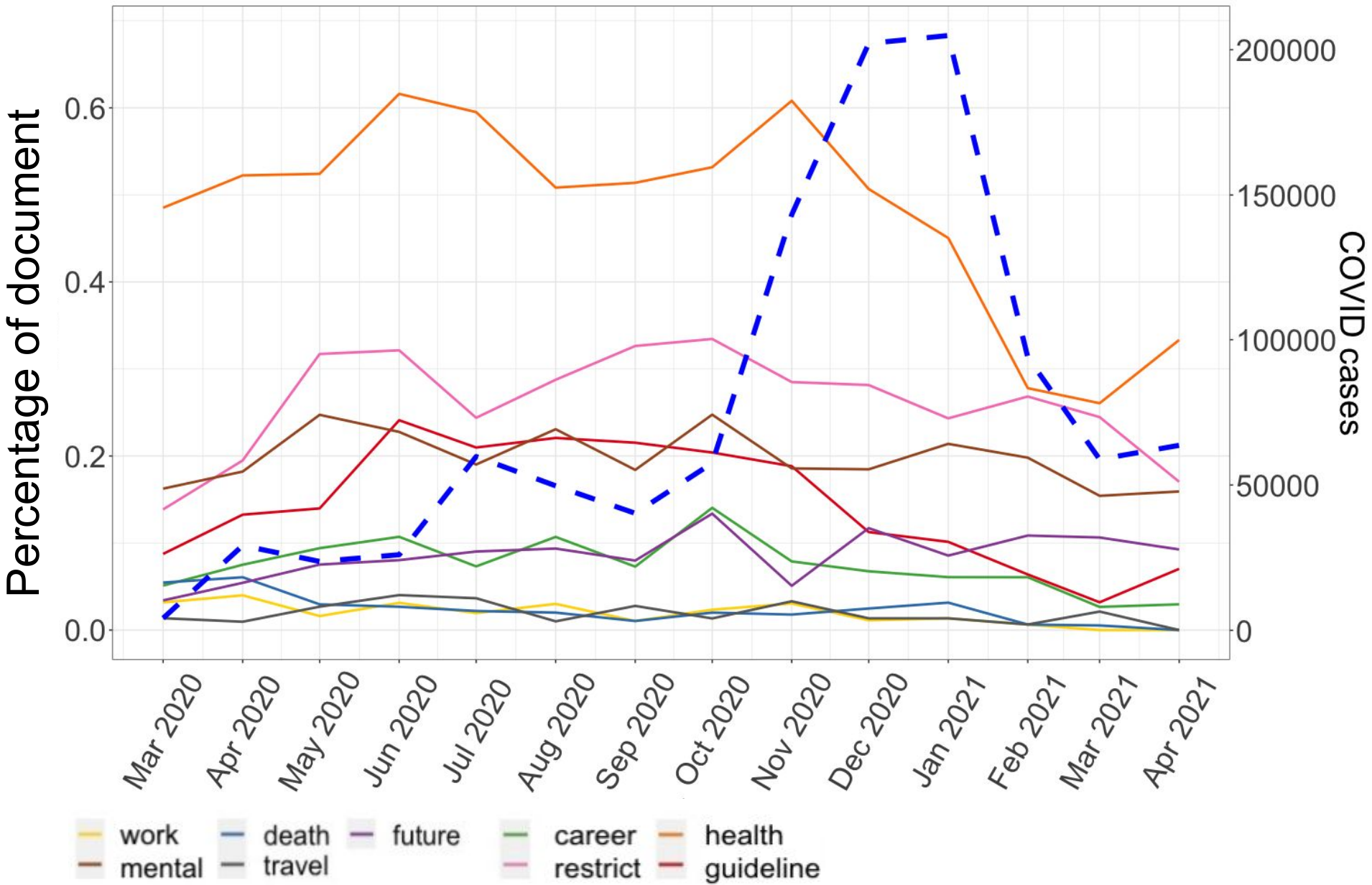}\\
    \caption{Changing Patterns of machine-annotated SOAs (Aggregated by Month). SOA: percentage of document mention specific SOA, the dashed blue line in the graph indicates COVID-19 cases reported in the U.S. } 
    \label{fig:content}
\end{figure}

Figure \ref{fig:content} presents the percentage of posts mentioning the nine SOAs for each study month using the SVM machine-annotated labels. To compare the anxiety index with reported COVID-19 cases, we retrieved COVID-19 case numbers from the website OurWorldinData \citep{ourworld}. According to website traffic tracked from Statista (\url{https://bit.ly/3DJf9wX}), Reddit traffic mainly comes from the U.S. The blue dotted line in Figure 2 indicates the number of COVID-19 cases reported in the U.S. We found that feeling anxious about infection risk (health risk) was the most prevalent concern throughout the study period, followed by struggling with COVID-19 restrictions (restrict).

\paragraph{Dominant Topics} Figure \ref{fig:content} shows that ``Health risk'' (worry about being infected with COVID-19) and ``COVID-19 restrictions'' (struggle with quarantine and social isolation) were mentioned most often in the Reddit posts throughout the time we examined. ``Health risk'' was mentioned most frequently in r/COVID19\_support during the entire period examined in this study (see Figure \ref{fig:content}), followed by ``COVID-19 restrictions''. Worries about ``Health risk'' remained high in the first eight months after the pandemic started, but it declined dramatically from the ninth month when there was a large wave of COVID-19 cases in Europe and the U.S. from October 2020 to January 2021, likely due to people visiting their loved ones during the holidays. 

\paragraph{Trends: Shift of Dominant Topic}
We observed a dramatic downward shift in worries ``Health risk'' and ``Disrespect for Health-related Guidelines'' as the timeline moved to 2021. The shift may be due to three reasons: 1) users' SOAs changed to topics that are not represented in our codings (see section: Model Validation 2021). 2) users' need for sharing or obtaining a specific type of information during the pandemic decreasing over time due to information fatigue \citet{skulmowski2021covid} or they had already obtained the information or coping strategies that they needed and were thus no longer seeking support on Reddit. 3) users' risk evaluations changed such that they no longer regarded Covid-19 risks as equally threatening.

\paragraph{Trends: Worry about Mental Health and the Future.} 
Although we see that worries about ``Health risk'', ``COVID-19 restrictions'' and ``Disrespect for Health-related Guidelines'' experienced a downward shift at the beginning of 2021, the number of posts mentioning ``mental health'' and ``future'' remained similar throughout the period we investigated. Unlike most of the SOAs that showed a downward trend at the beginning of 2021, feeling anxious about the future slowly but steadily increased during the pandemic. In the last month holdout model, the model performance of the ``mental health'' and ``future'' categories have been affected by the shift of SOAs, but the effect is less prominent compared with other categories such as ``COVID-19 restrictions'' and ``work risk''. This is likely due to the fact that mental health and future concerns are still mentioned very often in the 2021 posts.

\begin{table}[ht]
\begin{threeparttable}
\tiny
\begin{tabular}{p{0.3cm}p{0.4cm}p{0.5cm}p{0.5cm}p{0.5cm}p{0.5cm}p{0.45cm}p{0.5cm}p{0.35cm}p{0.35cm}} 
  \hline
   & intensity & finance & restrict & health & guide & work & mental & death & travel \\ 
  \hline
  
  finance & .11*&  &  &  &  &  &  &  &  \\ 
  restrict & .15*& -.04    &  &  &  &  &  &  &  \\ 
  health & .24*&  -.05     & -.12*     &  &  &  &  &  &  \\ 
  guide  & .14**&  .04   &  .04     &  .50** &  &  &  &  &  \\ 
  work  & .09$\cdot$&  .20** & -.11*  &  .28** &  .18** &  &  &  &  \\ 
  mental &.42** & .02     &  .21**   & -.10*     &  .03     & -.03     &  &  &  \\ 
  death & .24**  & -.04     & -.10*   &  .09*    &  -.02     &  .02     &  .12**   &  &  \\ 
  travel & .03 &  0.00     &  .06     &  0     & .01     & -.04     & -.10*    & -.07     &  \\ 
  future & 0.10$\cdot$ &  .19* &  .13** & -.20* & -.08$\cdot$     & -.09*     &  .16**    & -.02     & -.02     \\ 
   \hline
\end{tabular}
  
    \caption{Correlations Between machine-annotated SOAs. Pearson correlations, $\cdot$: $p < .05$, *: $p < .01$, **: $p<.001$, The p-values are corrected for multiple inference using Holm's method.} \label{tab:cor}
\end{threeparttable}
\end{table}

\subsection{How Did Anxiety-related Language Change Over Time?}
\begin{figure}
\centering
    \includegraphics[scale=0.35]{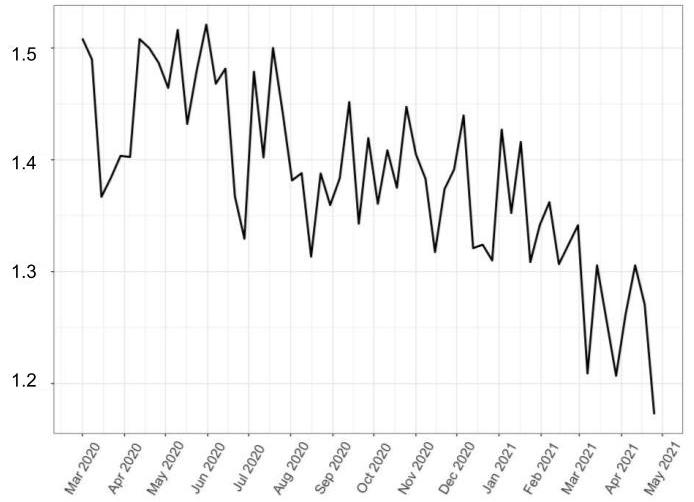}\\

    \caption{Intensity of machine-annotated anxiety-related language in COVID19 Support Dataset (Weekly).value: mean predicted value of anxiety-related language} 
    \label{fig:anxiety}
    
\end{figure}

Figure \ref{fig:anxiety} shows the weekly time series of the mean anxiety-related language intensity from machine annotations.  Users' anxiety-related language intensity took a downward trend about half a year after the pandemic started (August 2020). The decrease became more dramatic in February 2021, which correlates with several developed countries starting their vaccination programs (see Figure \ref{fig:content}). Note that unlike SOAs whose measurement quality shifted over time, our machine annotation on language intensity maintained good performance in the new incoming data (see Table \ref{tab:model}). 

Overall, we found that over time, Reddit users' language in describing the SOAs became less intense. This pattern may suggest that despite the pandemic situation not improving, users developed strategies to cope with anxiety issues after a while. Alternatively, they may have shared their worries in other spaces, such as in personal conversations instead of a Reddit forum.  

\subsubsection{Anxiety Language: High Intensity in Mental Health.} 

Since the language describing the SOAs reflects the intensity of these anxieties, we further conducted correlation analyses to understand which SOA was described with more intense anxiety-related language. Table \ref{tab:cor} shows that language intensity was mostly correlated with worries about deteriorating mental health ($r = 0.42$, $p < 0.001$), followed by worries about death ($r = 0.24$, $p < 0.001$) and becoming infected ($r = 0.24$, $p < 0.001$). Feeling anxious about the future was closely linked to feeling anxious about career advancement and COVID-19 restrictions.



\subsection{Feature Importance}

We ranked the coefficients of the linear kernel SVM models and extracted the top 10 important features for each classifier. Table \ref{tab:feaimp} shows the important features in each classifier.  In general, the top features for each classifier reflect the behavior or emotions we looked for during the annotation process. For example, the finance or career classifier included employment-related words and topics, which were slightly overlapped with the top features of the work risk classifier. Top  features for disrespecting health-related guidelines reflected disrespectful behaviors to the COVID19 guidelines (e.g., ``(not) wear mask'', ``social''). The anxiety-related language classifier had words indicating negative emotions ranked as top predictive features. 

\begin{table}[]
\begin{threeparttable}
\scriptsize
\begin{tabular}{p{1cm}|p{6.5cm}}
\hline
          & top 10 features  (ranked by coefficients)                                                                                        \\
          \hline
intensity &  sad, because, option, shopping, droplet, negemo(LIWC), younger, terrify, anxiety roof \\ 
finance   & job, lose job, hire, employment (LDA), work (LIWC), achieve(LIWC), lose, money (LIWC), future, reason why \\
restrict  & social, quarantine, toll, social activities (LDA), honest, uni, friend because,  proceed, know virus, sex  \\
guide     & plane, wear, destroy, wear mask, seriously, toronto, right thing, parties, request, refuse     \\
work      & come work, work, start work,  employee, job, homework, need money, blow, new job, apply                        \\
mental    & therapy, sane, depression, anxiety, silent, afford, alone, headline, everything, mental   \\
death     & dying, hospital, die, cancer, death (LIWC), need hospital, story, death, recover, need hear     \\
travel    & flight, fly, trip, plane, travel, florida, metro, airport, uber, hotel                           \\
future    & hopeless, vaccine, economy, absolute best, taken away, feel hopeless, end, oxford, forever, world \\
\hline
\end{tabular}

    \caption{Feature Importance. intensity: anxiety-related language intensity; LIWC: LIWC categories, LDA: LDA topic modeling.} \label{tab:feaimp}
\end{threeparttable}
\end{table}

\section{Discussion}
In this study, we use a thematic analysis to summarize the themes and sub-themes of aspects that trigger anxiety during a pandemic (see Figure \ref{fig:SOAgraph}). Then we used a computational approach to identify Reddit users' subjects of anxiety (SOAs) expressed in r/COVID-19\_support during the COVID-19 pandemic. To our knowledge, this is the first study that combines quantitative and qualitative analyses of social media data to provide insight into how users' focus of anxiety changed during the pandemic.

\subsection{Content Shift in Social Media Discussion}

Our trend analysis only focuses on categories that have reasonable F1 scores from the SVM model. We found that health risks, COVID-19 restrictions, and other people disrespecting health-related guidelines were the most prominent subjects that people worried about. Worries about health risks reduced dramatically starting Jan 2021, coinciding with when several countries, especially developed countries, started to plan vaccine rollouts. In our analysis of 2021 data, we found that worries about vaccine safety and efficacy had become dominant SOAs. On the one hand, we suspect that people may have been less anxious about infection risk due to vaccination programs. On the other hand, people may shift their major subject of anxiety from health risks of COVID-19 to vaccine safety.

Nevertheless, through our investigation on the last month hold-out set, we found that precision on ``disrespecting health-related guidelines'' was much lower on the October data compared with data from the previous month, which may imply that users' language surrounding anxiety towards people disrespecting health-related guidelines had changed. We also see a similar pattern in the evaluation of 2021 data (see Table \ref{tab:model}). We highlight the importance of updating our knowledge of how the textual content evolves while using social media text to construct automatic systems to infer people's mental health status.

\subsection{COVID-19 Restrictions Compromise Mental Health}

The percentage of posts mentioning mental health concerns remains similar even though there was a shift of topic as the timeline entered 2021. We found that posts worrying about COVID-19 restrictions were significantly correlated with users' mental health expressed in the same posts ($r = 0.21$, $p < 0.01$). In our thematic analysis, we found that people who worried about COVID-19 restrictions often felt insecure over how long and when to quarantine/distance, worried about meeting with friends, family and loved ones, or felt threatened by family members or roommates not sticking to quarantine/distancing regulations as closely as they did. Research is beginning to show that that prolonged isolation and limited physical space during isolation during the pandemic compromised mental health  \citep{pancani2021forced}. 

\subsection{Adjustment to the ``New Normal''}
We observe that users tended to use less intense language to describe the SOAs as the pandemic proceeded, even though the number of confirmed cases did not decrease. This may suggest that after coping with ongoing stressors for a long time, people may start to adapt to the new reality. Further, the sudden drop of anxiety-related language intensity around February 2021 correlates with the starting point of vaccination programs in the US and many European countries. Although speculative, this co-occurrence could indicate that vaccination programs may have reduced users' pandemic anxiety. 

\subsection{Application}
Annotation of social media posts is not meant to replace self-report measurement of anxiety levels or anxiety subjects. The types of information assessed with self-report measures do not necessarily exist in the social media text because Reddit users may be mindful of self-presentation as their posts are part of their larger ``Reddit identity''. 

However, the methods proposed in this paper can inform health providers about the mental health state of users in support-seeking discussion forums during a pandemic. Samples in our paper are not representative of social media platforms in general, but it is possible to adopt similar approaches to identify pandemic-related anxiety themes in other social media platforms and support forums. Health care providers can gauge the impact of the pandemic on the public's mental health and better prepare resources according to the estimated trends and widespread concerns expressed across various social media platforms and discussion forums.




\subsection{Limitations and Future Directions}

The thematic analysis for subjects of anxiety included only about ten posts per month (in total, 86). Thus, some of the subjects of anxiety may have been missed because ten posts only account for a small percentage of the posts in each month, especially when r/COVID19\_support became more popular in some months. Future studies can consider expanding the number of annotation examples in the initial stage for identifying themes, although such annotation is very resource expensive because annotation must be conducted with trained personnel with relevant background knowledge instead of relying on crowdsourcing. In addition, we found that a small percentage of posts (8\%) did not mention any SOAs and did not contain any anxiety language. As motivated in the Methods section, we did not construct separate classifiers to remove the noise from our dataset in this work. Future work should increase the annotated sample size and construct a classifier to remove posts only for information sharing.

Like in many other social media studies, users from social media platforms are not a representative sample of the general population. Many people choose not to browse or post on social media platforms for many reasons. The sample in this study may not be representative of social media users in general as well. Users from r/COVID19\_support may also have more heightened anxiety than the general population. Nevertheless, Reddit does not contain information about users' socio-demographic characteristics and refined analyses where we include control variables for relevant factors like users' age, gender, country, or ethnic group were therefore not possible. Survey studies examining the demographic characteristics of Reddit users estimated that around 90\% of users are under the age of 35, with a mean age of 25 years \citep{duggan20136}. We recognize that studying the mental health implications of this pandemic on all age groups - spanning the entire life course - is important and invite future studies to replicate these findings using younger and older samples.

Despite the above limitations, our method can be adopted to analyze subjects of anxiety and its language intensity across various social media user subgroups. By combining the results of many social media subgroups, we could gain a general idea of the subjects of anxiety among certain populations, especially young people in certain countries.

\section{Conclusion}
In this paper, we employed a computational approach to measure, describe, and understand what people worried about during the COVID-19 pandemic and how these patterns changed throughout the pandemic. We combined quantitative techniques with a content analysis approach to reveal the subjects of anxiety in a COVID-19 support community from Reddit. We found that oneself or a loved one becoming infected with COVID-19 was the most frequent source of anxiety and that feeling anxious about COVID-19 restrictions was frequently mentioned in conjunction with concerns about one's deteriorating mental health. We believe the present study contributes novel insights into the pandemic's impacts on population health. 



\bibliography{citation.bib}

\begin{thebibliography}{41}
\providecommand{\natexlab}[1]{#1}
\providecommand{\url}[1]{\texttt{#1}}
\providecommand{\urlprefix}{URL }
\expandafter\ifx\csname urlstyle\endcsname\relax
  \providecommand{\doi}[1]{doi:\discretionary{}{}{}#1}\else
  \providecommand{\doi}{doi:\discretionary{}{}{}\begingroup
  \urlstyle{rm}\Url}\fi

\bibitem[{Ahorsu et~al.(2020)Ahorsu, Lin, Imani, Saffari, Griffiths, and
  Pakpour}]{ahorsu2020fear}
Ahorsu, D.~K.; Lin, C.-Y.; Imani, V.; Saffari, M.; Griffiths, M.~D.; and
  Pakpour, A.~H. 2020.
\newblock The fear of COVID-19 scale: development and initial validation.
\newblock \emph{International journal of mental health and addiction} 1--9.

\bibitem[{Bargh, McKenna, and Fitzsimons(2002)}]{bargh2002can}
Bargh, J.~A.; McKenna, K.~Y.; and Fitzsimons, G.~M. 2002.
\newblock Can you see the real me? Activation and expression of the “true
  self” on the Internet.
\newblock \emph{Journal of social issues} 58(1): 33--48.

\bibitem[{Boe(2016)}]{praw}
Boe, B. 2016.
\newblock The Python Reddit API Wrapper 4.
\newblock \url{https://github.com/praw-dev/praw/}.
\newblock Accessed: 2021-08-31.

\bibitem[{Braun and Clarke(2006)}]{braun2006using}
Braun, V.; and Clarke, V. 2006.
\newblock Using thematic analysis in psychology.
\newblock \emph{Qualitative research in psychology} 3(2): 77--101.

\bibitem[{Britten(1995)}]{britten1995qualitative}
Britten, N. 1995.
\newblock Qualitative research: qualitative interviews in medical research.
\newblock \emph{Bmj} 311(6999): 251--253.

\bibitem[{Brown et~al.(2003)Brown, Nesse, Vinokur, and
  Smith}]{brown2003providing}
Brown, S.~L.; Nesse, R.~M.; Vinokur, A.~D.; and Smith, D.~M. 2003.
\newblock Providing social support may be more beneficial than receiving it:
  Results from a prospective study of mortality.
\newblock \emph{Psychological science} 14(4): 320--327.

\bibitem[{Chancellor and De~Choudhury(2020)}]{chancellor2020methods}
Chancellor, S.; and De~Choudhury, M. 2020.
\newblock Methods in predictive techniques for mental health status on social
  media: a critical review.
\newblock \emph{NPJ digital medicine} 3(1): 1--11.

\bibitem[{Chen(2021)}]{chen2021monitoring}
Chen, L. 2021.
\newblock \emph{Monitoring depressive symptoms using social media data}.
\newblock {PhD} dissertation, The University of Edinburgh.

\bibitem[{Chen et~al.(2020)Chen, Magdy, Whalley, and
  Wolters}]{chen2020examining}
Chen, L.; Magdy, W.; Whalley, H.; and Wolters, M.~K. 2020.
\newblock Examining the role of mood patterns in predicting self-reported
  depressive symptoms.
\newblock In \emph{12th ACM Conference on Web Science}, 164--173.

\bibitem[{Coleman and Liau(1975)}]{coleman1975computer}
Coleman, M.; and Liau, T.~L. 1975.
\newblock A computer readability formula designed for machine scoring.
\newblock \emph{Journal of Applied Psychology} 60(2): 283.

\bibitem[{Conway~III, Woodard, and Zubrod(2020)}]{conway2020social}
Conway~III, L.~G.; Woodard, S.~R.; and Zubrod, A. 2020.
\newblock Social psychological measurements of COVID-19: Coronavirus perceived
  threat, government response, impacts, and experiences questionnaires .

\bibitem[{Duggan and Smith(2013)}]{duggan20136}
Duggan, M.; and Smith, A. 2013.
\newblock 6\% of online adults are reddit users.
\newblock \emph{Pew Internet \& American Life Project} 3: 1--10.

\bibitem[{Guntuku et~al.(2019)Guntuku, Preotiuc-Pietro, Eichstaedt, and
  Ungar}]{guntuku2019twitter}
Guntuku, S.~C.; Preotiuc-Pietro, D.; Eichstaedt, J.~C.; and Ungar, L.~H. 2019.
\newblock What twitter profile and posted images reveal about depression and
  anxiety.
\newblock In \emph{Proceedings of the International AAAI Conference on Web and
  Social Media}, volume~13, 236--246.

\bibitem[{Guntuku et~al.(2020)Guntuku, Sherman, Stokes, Agarwal, Seltzer,
  Merchant, and Ungar}]{guntuku2020tracking}
Guntuku, S.~C.; Sherman, G.; Stokes, D.~C.; Agarwal, A.~K.; Seltzer, E.;
  Merchant, R.~M.; and Ungar, L.~H. 2020.
\newblock Tracking mental health and symptom mentions on twitter during
  covid-19.
\newblock \emph{Journal of general internal medicine} 35(9): 2798--2800.

\bibitem[{Huang and Zhao(2020)}]{huang2020generalized}
Huang, Y.; and Zhao, N. 2020.
\newblock Generalized anxiety disorder, depressive symptoms and sleep quality
  during COVID-19 outbreak in China: a web-based cross-sectional survey.
\newblock \emph{Psychiatry research} 112954.

\bibitem[{Hutto and Gilbert(2014)}]{hutto2014vader}
Hutto, C.; and Gilbert, E. 2014.
\newblock Vader: A parsimonious rule-based model for sentiment analysis of
  social media text.
\newblock In \emph{Proceedings of the International AAAI Conference on Web and
  Social Media}, volume~8.

\bibitem[{Jones and Silver(2020)}]{jones2020not}
Jones, N.~M.; and Silver, R.~C. 2020.
\newblock This is not a drill: Anxiety on Twitter following the 2018 Hawaii
  false missile alert.
\newblock \emph{American Psychologist} 75(5): 683.

\bibitem[{Liaw et~al.(2018)Liaw, Liang, Nishihara, Moritz, Gonzalez, and
  Stoica}]{liaw2018tune}
Liaw, R.; Liang, E.; Nishihara, R.; Moritz, P.; Gonzalez, J.~E.; and Stoica, I.
  2018.
\newblock Tune: A Research Platform for Distributed Model Selection and
  Training.
\newblock \emph{arXiv preprint arXiv:1807.05118} .

\bibitem[{Low et~al.(2020)Low, Rumker, Talkar, Torous, Cecchi, and
  Ghosh}]{low2020natural}
Low, D.~M.; Rumker, L.; Talkar, T.; Torous, J.; Cecchi, G.; and Ghosh, S. 2020.
\newblock Natural language processing reveals vulnerable mental health support
  groups and heightened health anxiety on Reddit during COVID-19. .

\bibitem[{{Max Roxer and colleagues}(2021)}]{ourworld}
{Max Roxer and colleagues}. 2021.
\newblock Coronavirus Pandemic (COVID-19) – the data.
\newblock \url{https://ourworldindata.org/coronavirus-data}.
\newblock Accessed: 2021-05-13.

\bibitem[{Maxwell et~al.(2020)Maxwell, Robinson, Williams, and
  Keaton}]{maxwell2020short}
Maxwell, D.; Robinson, S.~R.; Williams, J.~R.; and Keaton, C. 2020.
\newblock " A Short Story of a Lonely Guy": A Qualitative Thematic Analysis of
  Involuntary Celibacy Using Reddit.
\newblock \emph{Sexuality \& Culture} 24(6).

\bibitem[{McElroy et~al.(2020)McElroy, Patalay, Moltrecht, Shevlin, Shum,
  Creswell, and Waite}]{mcelroy2020demographic}
McElroy, E.; Patalay, P.; Moltrecht, B.; Shevlin, M.; Shum, A.; Creswell, C.;
  and Waite, P. 2020.
\newblock Demographic and health factors associated with pandemic anxiety in
  the context of COVID-19.
\newblock \emph{British Journal of Health Psychology} 25(4): 934--944.

\bibitem[{Mimno et~al.(2011)Mimno, Wallach, Talley, Leenders, and
  McCallum}]{mimno2011optimizing}
Mimno, D.; Wallach, H.; Talley, E.; Leenders, M.; and McCallum, A. 2011.
\newblock Optimizing semantic coherence in topic models.
\newblock In \emph{Proceedings of the 2011 conference on empirical methods in
  natural language processing}, 262--272.

\bibitem[{Murnane and Counts(2014)}]{murnane2014unraveling}
Murnane, E.~L.; and Counts, S. 2014.
\newblock Unraveling abstinence and relapse: smoking cessation reflected in
  social media.
\newblock In \emph{Proceedings of the SIGCHI conference on human factors in
  computing systems}, 1345--1354.

\bibitem[{Nik{\v{c}}evi{\'c} and Spada(2020)}]{nikvcevic2020covid}
Nik{\v{c}}evi{\'c}, A.~V.; and Spada, M.~M. 2020.
\newblock The COVID-19 Anxiety Syndrome Scale: development and psychometric
  properties.
\newblock \emph{Psychiatry research} 292: 113322.

\bibitem[{Pancani et~al.(2021)Pancani, Marinucci, Aureli, and
  Riva}]{pancani2021forced}
Pancani, L.; Marinucci, M.; Aureli, N.; and Riva, P. 2021.
\newblock Forced Social Isolation and Mental Health: A Study on 1,006 Italians
  Under COVID-19 Lockdown.
\newblock \emph{Frontiers in Psychology} 12: 1540.

\bibitem[{Pedregosa et~al.(2011)Pedregosa, Varoquaux, Gramfort, Michel,
  Thirion, Grisel, Blondel, Prettenhofer, Weiss, Dubourg, Vanderplas, Passos,
  Cournapeau, Brucher, Perrot, and Duchesnay}]{scikit-learn}
Pedregosa, F.; Varoquaux, G.; Gramfort, A.; Michel, V.; Thirion, B.; Grisel,
  O.; Blondel, M.; Prettenhofer, P.; Weiss, R.; Dubourg, V.; Vanderplas, J.;
  Passos, A.; Cournapeau, D.; Brucher, M.; Perrot, M.; and Duchesnay, E. 2011.
\newblock Scikit-learn: Machine Learning in {P}ython.
\newblock \emph{Journal of Machine Learning Research} 12: 2825--2830.

\bibitem[{Pennebaker et~al.(2015)Pennebaker, Boyd, Jordan, and
  Blackburn}]{pennebaker2015development}
Pennebaker, J.~W.; Boyd, R.~L.; Jordan, K.; and Blackburn, K. 2015.
\newblock The development and psychometric properties of LIWC2015.
\newblock Technical report.

\bibitem[{Pfefferbaum and North(2020)}]{pfefferbaum2020mental}
Pfefferbaum, B.; and North, C.~S. 2020.
\newblock Mental health and the Covid-19 pandemic.
\newblock \emph{New England Journal of Medicine} 383(6): 510--512.

\bibitem[{Roohafza et~al.(2014)Roohafza, Afshar, Keshteli, Mohammadi, Feizi,
  Taslimi, and Adibi}]{roohafza2014s}
Roohafza, H.~R.; Afshar, H.; Keshteli, A.~H.; Mohammadi, N.; Feizi, A.;
  Taslimi, M.; and Adibi, P. 2014.
\newblock What's the role of perceived social support and coping styles in
  depression and anxiety?
\newblock \emph{Journal of research in medical sciences: the official journal
  of Isfahan University of Medical Sciences} 19(10): 944.

\bibitem[{Roy et~al.(2020)Roy, Tripathy, Kar, Sharma, Verma, and
  Kaushal}]{roy2020study}
Roy, D.; Tripathy, S.; Kar, S.~K.; Sharma, N.; Verma, S.~K.; and Kaushal, V.
  2020.
\newblock Study of knowledge, attitude, anxiety \& perceived mental healthcare
  need in Indian population during COVID-19 pandemic.
\newblock \emph{Asian Journal of Psychiatry} 102083.

\bibitem[{Sanh et~al.(2019)Sanh, Debut, Chaumond, and
  Wolf}]{sanh2019distilbert}
Sanh, V.; Debut, L.; Chaumond, J.; and Wolf, T. 2019.
\newblock DistilBERT, a distilled version of BERT: smaller, faster, cheaper and
  lighter.
\newblock \emph{arXiv preprint arXiv:1910.01108} .

\bibitem[{Shen and Rudzicz(2017)}]{shen2017detecting}
Shen, J.~H.; and Rudzicz, F. 2017.
\newblock Detecting anxiety through reddit.
\newblock In \emph{Proceedings of the Fourth Workshop on Computational
  Linguistics and Clinical Psychology—From Linguistic Signal to Clinical
  Reality}, 58--65.

\bibitem[{Shen et~al.(2018)Shen, Jia, Shen, Feng, He, Luan, Tang, Tiropanis,
  Chua, and Hall}]{shen2018cross}
Shen, T.; Jia, J.; Shen, G.; Feng, F.; He, X.; Luan, H.; Tang, J.; Tiropanis,
  T.; Chua, T.~S.; and Hall, W. 2018.
\newblock Cross-domain depression detection via harvesting social media.
\newblock International Joint Conferences on Artificial Intelligence.

\bibitem[{Skulmowski and Standl(2021)}]{skulmowski2021covid}
Skulmowski, A.; and Standl, B. 2021.
\newblock COVID-19 information fatigue? A case study of a German university
  website during two waves of the pandemic.
\newblock \emph{Human Behavior and Emerging Technologies} .

\bibitem[{Smith et~al.(2020)Smith, Jacob, Yakkundi, McDermott, Armstrong,
  Barnett, L{\'o}pez-S{\'a}nchez, Martin, Butler, and
  Tully}]{smith2020correlates}
Smith, L.; Jacob, L.; Yakkundi, A.; McDermott, D.; Armstrong, N.~C.; Barnett,
  Y.; L{\'o}pez-S{\'a}nchez, G.~F.; Martin, S.; Butler, L.; and Tully, M.~A.
  2020.
\newblock Correlates of symptoms of anxiety and depression and mental wellbeing
  associated with COVID-19: a cross-sectional study of UK-based respondents.
\newblock \emph{Psychiatry research} 291: 113138.

\bibitem[{Taylor et~al.(2020)Taylor, Landry, Paluszek, Fergus, McKay, and
  Asmundson}]{taylor2020development}
Taylor, S.; Landry, C.~A.; Paluszek, M.~M.; Fergus, T.~A.; McKay, D.; and
  Asmundson, G.~J. 2020.
\newblock Development and initial validation of the COVID Stress Scales.
\newblock \emph{Journal of Anxiety Disorders} 72: 102232.

\bibitem[{Wiederhold(2020)}]{wiederhold2020using}
Wiederhold, B.~K. 2020.
\newblock Using social media to our advantage: alleviating anxiety during a
  pandemic.
\newblock \emph{Cyberpsychology, Behavior, and Social Networking} 23(4):
  197--198.

\bibitem[{Wilkinson(2001)}]{wilkinson2001anxiety}
Wilkinson, I. 2001.
\newblock \emph{Anxiety in a risk society}.
\newblock Psychology Press.

\bibitem[{Wolf et~al.(2019)Wolf, Debut, Sanh, Chaumond, Delangue, Moi, Cistac,
  Rault, Louf, Funtowicz et~al.}]{wolf2019huggingface}
Wolf, T.; Debut, L.; Sanh, V.; Chaumond, J.; Delangue, C.; Moi, A.; Cistac, P.;
  Rault, T.; Louf, R.; Funtowicz, M.; et~al. 2019.
\newblock Huggingface's transformers: State-of-the-art natural language
  processing.
\newblock \emph{arXiv preprint arXiv:1910.03771} .

\bibitem[{Yap, Wright, and Jorm(2011)}]{yap2011influence}
Yap, M.~B.; Wright, A.; and Jorm, A.~F. 2011.
\newblock The influence of stigma on young people’s help-seeking intentions
  and beliefs about the helpfulness of various sources of help.
\newblock \emph{Social Psychiatry and Psychiatric Epidemiology} 46(12):
  1257--1265.

\end{thebibliography}

\appendix

\begin{table*}[!htb]
\begin{threeparttable}
\caption{Comparing COVID-19 anxiety measurement scale with SOA annotation. This table shows that our SOAs annotation overlaps with many items in multiple COVID-19 anxiety measurement scales. Scale-item: items in the COVID-19 scales. SOA: SOA in our annotation task. SOAs annotation guidelines: annotation guidelines for subjects of anxiety.}
\small
\begin{tabular}{{p{6cm}|p{2cm}|p{8cm}}}
\hline
Scale-item      &  SOA      &   \textbf{SOAs annotation guidelines} \\

\hline
 \multicolumn{2}{c}{\textbf{The Coronavirus Anxiety Syndrome Scale (CASS) -18 items}}                                                    \\
 
 \hline

Avoided using public transport                                                                                                                          & travel           & Author expresses feeling anxious about being infected during the course of traveling, taking trains, flights or public transportation. Feeling anxious about how to travel when the city is in lockdown and frustrated about not being able to travel.                 \\
\hline

Checked myself for symptoms of coronavirus (COVID-19).                  & health, infected & Author shows signs of fear or anxious about self, a friend/family member will be infected with COVID-19. Or redditor listed symptoms and asked whether they are infected. Or redditor listed symptoms and thought they were infected.                                                                                                          \\
\hline

Concerned about not having adhered strictly to social distancing guidelines for coronavirus (COVID-19).                                                 & guideline        & Author shows signs of fear or anxious that family members, friends and strangers do not follow preventative measurements in the supermarket, public transportation and any other enclosed environment, feeling anxious about having to use a face mask forever. Feeling anxious about what should be the appropriate preventative measure in a given situation. \\\hline

Researched symptoms of coronavirus (COVID-19) at the cost of off-line social activities such as spending time with friends/family.                      & health, infected & Author shows signs of fear or feeling anxious that self, a friend or family member will be infected with COVID-19. Post author researched and listed symptoms, asking whether they are infected.                                                                                                                                                         \\\hline

Paying close attention to others displaying possible symptoms of coronavirus (COVID-19).                                                                & health, infected & Author shows signs of fear or anxiety that self, a friend or family member will be infected with COVID-19. users researched and listed symptoms, asking whether they are infected.                                                                                                                                                         \\\hline
Imagined what could happen to my family members if they contracted coronavirus (COVID-19).                                                              & death            & Author shows signs of fear or anxiety about that self, a friend, family member or someone they know will die of COVID-19 or become critically ill.                                                                                                                                                                                                                                 \\\hline
 \multicolumn{2}{c}{\textbf{COVID-19 Concern Questionnaire - 6 items}}       \\\hline
Worried I or loved ones will get COVID-19.                                                                                                              &   death               &  Author shows signs of fear or anxiety about that self, a friend, family member, or someone they know will die of COVID-19 or become critically ill.                                                                                                                                                                                                                                                                                                                                     \\\hline
Practiced social distancing due to anxiety about getting COVID-19.                                                                                        &    quarantine, social distancing and other COVID-19 restrictions            &                                                                                                                                             Author shows anxiety or fear about COVID-19 restrictions, such as when to quarantine, must quarantine, doesn't know when quarantine will finish, or conflicts with family members in a quarantine situation. Struggles with quarantine and not seeing loved ones.                                                                                                                                                                                         \\\hline
Felt threatened, fear of, anxious about COVID-19.                                                                                                               &     death             &    Author shows signs of fear or anxiety that self, a friend, family member, or someone they know will die of COVID-19 or become critically ill.                                                                                                                                                                                                                                                                                                                                                                                                                      \\\hline
 \multicolumn{2}{c}{\textbf{COVID-19 Experiences Questionnaire - 14 items}}                                                                                                                                                                         \\\hline
Showed COVID-19 symptoms.                                                                                                                               &     health risk             &    Author shows signs of fear or anxiety  about being infected with COVID-19. * Author lists symptoms and asks whether they are infected. * Author lists symptoms and thinks they are infected * Author was in direct contact with people who tested positive and is now afraid of being infected. * Author went to public places or travelled with public transport and is afraid of infection.

                                                                                                                                 \\\hline
Lost job-related income, financial insecurity.                                                                                                          &  financial/career                 &            Author expresses fear or anxiety about their financial, academic, or career situation. * Consider if the Author is seeking financial/ career advice. * Consider if the Author is anxious about COVID-19 interrupting school or career advancement.                                                                                                                                                                                                                                                                                                                           \\\hline
Depressed, negative impact on psychological health.                                                                                                     &     mental health             &                                                                                                       Author shows signs of fear or anxiety  about the mental health of self, friends or family. Author focuses on expressing their anxiety. Author seems to struggle with mental health issues. \\\hline
     
\hline

\end{tabular}

\label{tab:SOWanno}
\end{threeparttable}
\end{table*}

\end{document}